\documentclass[twocolumn,epjc3upd]{svjour3}          
\RequirePackage[T1]{fontenc}
\smartqed  
\RequirePackage{graphicx}
\RequirePackage{mathptmx}      
\RequirePackage{flushend}
\RequirePackage[numbers,sort&compress]{natbib}
\RequirePackage[colorlinks,citecolor=blue,urlcolor=blue,linkcolor=blue]{hyperref}
\usepackage{arydshln}
\journalname{Eur. Phys. J. C}
\usepackage[switch]{lineno}
\usepackage{upgreek}
\newcommand{\unit}[2]{#1\,{\rm #2}}
\newcommand{\Rn}{$^{222}$Rn}
\newcommand{\Ra}{$^{226}$Ra}

\begin{document}

\title{\Rn\ emanation measurements for the XENON1T experiment}

\author{E.~Aprile\thanksref{addr3}
\and
J.~Aalbers\thanksref{addr13}
\and
F.~Agostini\thanksref{addr0}
\and
M.~Alfonsi\thanksref{addr5}
\and
L.~Althueser\thanksref{addr7}
\and
F.~D.~Amaro\thanksref{addr2}
\and
V.~C.~Antochi\thanksref{addr13}
\and
E.~Angelino\thanksref{addr15}
\and
J.~R.~Angevaare\thanksref{addr8}
\and
F.~Arneodo\thanksref{addr9}
% ...
(XENON Collaboration\thanksref{email1}). }
\newcommand{\bologna}{Department of Physics and Astronomy, University of Bologna and INFN-Bologna, 40126 Bologna, Italy}
\newcommand{\chicago}{Department of Physics \& Kavli Institute for Cosmological Physics, University of Chicago, Chicago, IL 60637, USA}
\newcommand{\coimbra}{LIBPhys, Department of Physics, University of Coimbra, 3004-516 Coimbra, Portugal}
\newcommand{\columbia}{Physics Department, Columbia University, New York, NY 10027, USA}
\newcommand{\lngs}{INFN-Laboratori Nazionali del Gran Sasso and Gran Sasso Science Institute, 67100 L'Aquila, Italy}
\newcommand{\mainz}{Institut f\"ur Physik \& Exzellenzcluster PRISMA, Johannes Gutenberg-Universit\"at Mainz, 55099 Mainz, Germany}
\newcommand{\heidelberg}{Max-Planck-Institut f\"ur Kernphysik, 69117 Heidelberg, Germany}
\newcommand{\munster}{Institut f\"ur Kernphysik, Westf\"alische Wilhelms-Universit\"at M\"unster, 48149 M\"unster, Germany}
\newcommand{\nikhef}{Nikhef and the University of Amsterdam, Science Park, 1098XG Amsterdam, Netherlands}
\newcommand{\nyuad}{New York University Abu Dhabi, Abu Dhabi, United Arab Emirates}
\newcommand{\purdue}{Department of Physics and Astronomy, Purdue University, West Lafayette, IN 47907, USA}
\newcommand{\rpi}{Department of Physics, Applied Physics and Astronomy, Rensselaer Polytechnic Institute, Troy, NY 12180, USA}
\newcommand{\rice}{Department of Physics and Astronomy, Rice University, Houston, TX 77005, USA}
\newcommand{\stockholm}{Oskar Klein Centre, Department of Physics, Stockholm University, AlbaNova, Stockholm SE-10691, Sweden}
\newcommand{\subatech}{SUBATECH, IMT Atlantique, CNRS/IN2P3, Universit\'e de Nantes, Nantes 44307, France}
\newcommand{\torino}{INAF-Astrophysical Observatory of Torino, Department of Physics, University  of  Torino and  INFN-Torino,  10125  Torino,  Italy}
\newcommand{\ucsd}{Department of Physics, University of California San Diego, La Jolla, CA 92093, USA}
\newcommand{\wis}{Department of Particle Physics and Astrophysics, Weizmann Institute of Science, Rehovot 7610001, Israel}
\newcommand{\zurich}{Physik-Institut, University of Z\"urich, 8057  Z\"urich, Switzerland}
\newcommand{\paris}{LPNHE, Sorbonne Universit\'{e}, Universit\'{e} de Paris, CNRS/IN2P3, Paris, France}
\newcommand{\freiburg}{Physikalisches Institut, Universit\"at Freiburg, 79104 Freiburg, Germany}
\newcommand{\lal}{Universit\'{e} Paris-Saclay, CNRS/IN2P3, IJCLab, 91405 Orsay, France}
\newcommand{\napels}{Department of Physics ``Ettore Pancini'', University of Napoli and INFN-Napoli, 80126 Napoli, Italy}
\newcommand{\nagoya}{Kobayashi-Maskawa Institute for the Origin of Particles and the Universe, and Institute for Space-Earth Environmental Research, Nagoya University, Furo-cho, Chikusa-ku, Nagoya, Aichi 464-8602, Japan}
\newcommand{\laquila}{Department of Physics and Chemistry, University of L'Aquila, 67100 L'Aquila, Italy}
\newcommand{\tokyo}{Kamioka Observatory, Institute for Cosmic Ray Research, and Kavli Institute for the Physics and Mathematics of the Universe (WPI), the University of Tokyo, Higashi-Mozumi, Kamioka, Hida, Gifu 506-1205, Japan}
\newcommand{\kobe}{Department of Physics, Kobe University, Kobe, Hyogo 657-8501, Japan}
\newcommand{\ucla}{Physics \& Astronomy Department, University of California, Los Angeles, CA 90095, USA}
\newcommand{\alsoatferrara}{INFN, Sez. di Ferrara and Dip. di Fisica e Scienze della Terra, Universit\`a di Ferrara, via G. Saragat 1, Edificio C, I-44122 Ferrara (FE), Italy}
\newcommand{\alsoatsuny}{Simons Center for Geometry and Physics and C. N. Yang Institute for Theoretical Physics, SUNY, Stony Brook, NY, USA}
\newcommand{\alsoatutrecht}{Institute for Subatomic Physics, Utrecht University, Utrecht, Netherlands}
\newcommand{\alsoatcoimbrapoli}{Coimbra Polytechnic - ISEC, Coimbra, Portugal}
\newcommand{\alsoatiarnagoya}{Institute for Advanced Research, Nagoya University, Nagoya, Aichi 464-8601, Japan}
\authorrunning{XENON Collaboration}
\thankstext{addr29}{Also at \alsoatsuny}
\thankstext{addr30}{Also at \alsoatutrecht}
\thankstext{addr32}{Also at \alsoatiarnagoya}
\thankstext{addr31}{Also at \alsoatcoimbrapoli}
\thankstext{addr28}{Also at \alsoatferrara}

\thankstext{email3}{\texttt{h.simgen@mpi-hd.mpg.de}}
\thankstext{email2}{\texttt{natascha.rupp@mpi-hd.mpg.de}}
\thankstext{email1}{\texttt{xenon@lngs.infn.it}}

\institute{\columbia\label{addr3}
\and
\stockholm\label{addr13}
\and
\bologna\label{addr0}
\and
\mainz\label{addr5}
\and
\munster\label{addr7}
\and
\coimbra\label{addr2}
\and
\torino\label{addr15}
\and
\nikhef\label{addr8}
\and
\nyuad\label{addr9}
\and
\zurich\label{addr18}
\and
\rpi\label{addr11}
\and
\wis\label{addr17}
\and
\heidelberg\label{addr6}
\and
\napels\label{addr22}
\and
\purdue\label{addr10}
\and
\freiburg\label{addr20}
\and
\subatech\label{addr14}
\and
\laquila\label{addr24}
\and
\lngs\label{addr4}
\and
\paris\label{addr19}
\and
\chicago\label{addr1}
\and
\tokyo\label{addr25}
\and
\nagoya\label{addr23}
\and
\lal\label{addr21}
\and
\kobe\label{addr26}
\and
\ucsd\label{addr16}
\and
\rice\label{addr12}
\and
\ucla\label{addr27}
}

\maketitle

\begin{abstract}

The selection of low-radioactive construction materials is of utmost importance for the success of low-energy rare event search experiments. Besides radioactive contaminants in the bulk, the emanation of radioactive radon atoms from material surfaces attains increasing relevance in the effort to further reduce the background of such experiments. In this work, we present the \Rn\ emanation measurements performed for the XENON1T dark matter experiment. Together with the bulk impurity screening campaign, the results enabled us to select the radio-purest construction materials, targeting a \Rn\ activity concentration of $\unit{10}{\upmu Bq/kg}$ in $\unit{3.2}{t}$ of xenon. The knowledge of the distribution of the \Rn\ sources allowed us to selectively eliminate problematic components in the course of the experiment. The predictions from the emanation measurements were compared to data of the \Rn\ activity concentration in XENON1T. The final \Rn\ activity concentration of $\unit{(4.5\pm0.1)}{\upmu Bq/kg}$ in the target of XENON1T is the lowest ever achieved in a xenon dark matter experiment.

\end{abstract}

\section{Introduction} 

Many cosmological observations suggest that a large fraction of the total matter density of the Universe is made up of an unknown form of dark matter \cite{planck_cosmo_constrains}. However, despite a large experimental effort, dark matter has not yet been discovered. XENON1T \cite{instrument_xenon} was the largest and most sensitive so far in the series of XENON direct dark matter search experiments \cite{Xe10, Xe100}. Its successor XENONnT will start data-taking in 2020. The primary aim of these experiments is the detection of Weakly Interacting Massive Particles (WIMPs), a promising dark matter candidate \cite{strigari}. As in other astroparticle physics experiments \cite{nexo,LZ,pandaX} liquid xenon is used as an efficient target for particle detection. The XENON detectors are dual-phase time projection chambers (TPCs) with a gaseous layer of xenon on top of the target. Particles interacting with xenon nuclei or the atomic electrons create scintillation light and ionization electrons. The light is detected by two arrays of UV-sensitive photomultiplier tubes (PMTs) on top and bottom of the detector. The ionization electrons are drifted upwards to the liquid-gas interface, where they are extracted and create a second, delayed scintillation light signal. Both signals are used to gain information about the location and energy of the interaction. They are also used to reject background either by the event multiplicity (multi-scatter versus single-scatter events) or by the type of interaction (electronic recoil versus nuclear recoil events).

XENON1T operated for two years, starting from December 2016. Similarly to other astroparticle physics experiment looking for rare events, it required an extremely low background level. Throughout the different generations of the XENON experiments, external background sources have been suppressed, e.g. by an improved external shield and xenon self-shielding and by the mitigation of radioactivity from materials. Their level was marginal in XENON1T and intrinsic background sources became dominant. Among them, the radioactive isotope $^{222}$Rn induced the leading background component \cite{xenon_physics_reach}. Its long-lived mother nucleus $^{226}$Ra (T$_{1/2} =  \unit{1600}{years}$) is part of the primordial $^{238}$U decay chain and thus present in most materials. Once $^{226}$Ra decays, the created noble gas isotope $^{222}$Rn may emanate from inner surfaces into the xenon volume.

As $^{222}$Rn has a relatively long half-life (T$_{1/2} =  \unit{3.8}{days}$), it can reach the active dark matter target, where the $\upbeta$-decays of its daughter isotope $^{214}$Pb can mimic signal events.  To achieve the scientific goal of XENON1T, a $^{222}$Rn activity concentration of $\unit{10}{\upmu Bq/kg}$ was required \cite{xenon_physics_reach}. Other radon isotopes may also lead to background events. However, their contribution was strongly suppressed due to their small abundance in the detector and much shorter half-lives, that did not allow for their dispersion within the target volume. \linebreak XENON1T has performed a comprehensive bulk impurity screening campaign to select radio-pure materials using High Purity Germanium (HPGe) spectroscopy and Inductively \linebreak Coupled Plasma Mass Spectrometry (ICP-MS) \cite{gamma_screening_paper}.  However, the measured $^{226}$Ra bulk activity can in general not be used to predict how much $^{222}$Rn emanates from the material, because surface impurities may become dominant. This made dedicated $^{222}$Rn emanation measurements necessary, which are described in this article.

There are two ways for a radon atom to leave the material in which it is produced: by recoil or by diffusion \cite{balek}. In the first case, the decay occurs directly on or below the material's surface. The recoil energy, which the radon nucleus receives during the $\upalpha$-decay of its mother radium nuclide ($\unit{85}{keV}$ \cite{recoil-reference} in the case of \Rn), is sufficient to eject it from the material. In the second case, the radon atom diffuses inside the bulk of the material. If it reaches a boundary surface before its decay, it will be emanated. Data on radon diffusion in metals are rare, but its diffusion coefficient is even smaller than the minuscule one of xenon in metals \cite{xe-diffusion}. Thus, it is reasonable to assume that radon diffusion plays a significant role only in soft or porous materials. As a consequence, a radon emanation measurement of metals is mostly a test of surface impurities. 

The article is structured as follows. Section \ref{ema-technique} discusses the employed \Rn\ assay techniques. In section \ref{individual_sample}, we pre\-sent the screening results of investigated materials, most of which were eventually used for the construction of \linebreak XENON1T. Section \ref{sub_system} describes the overall \Rn\ emanation measurements of the assembled XENON1T detector and gives a summary on the inferred $^{222}$Rn budget, as well as a comparison to XENON1T data. We also describe radon reduction methods that were applied during detector operation. We close with a summary and outlook in section \ref{summary}.

\section{The \Rn\ assay techniques}
\label{ema-technique}

For XENON1T we used two methods to study \Rn\ emanation. The most sensitive method applied ultra-low background miniaturized proportional counters \cite{GERDA_welds}, developed for the GALLEX solar neutrino experiment \cite{Wink-paper}. These devices are made of high purity synthetic quartz with an iron cathode and a thin tungsten-wire ($13~\upmu$m diameter) anode. The active volume of the counters is around $\unit{1}{cm^3}$ and the counting gas consists of $\unit{90}{\%}$ argon and $\unit{10}{\%}$ methane to which the radon to be measured is added. \Rn\ atoms decay by $\upalpha$-disintegration, followed by two further $\upalpha$-decays from $^{218}$Po and $^{214}$Po until the long-lived $^{210}$Pb breaks the secular equilibrium. Cosmic muons as well as environmental $\upbeta$- and $\upgamma$-radiation cannot deposit energies above $\unit{50}{keV}$ in the miniaturized counters. In contrast, $\upalpha$-decays exhibit a larger energy deposition, allowing for their clear identification. The detection efficiency for the three $\upalpha$-decays is not equal, because the gaseous \Rn\ is distributed in the entire counter volume, while polonium ions deposit on surfaces. On average ($49.3 \pm 2.0$)\,\% of all $\upalpha$-decays are detected, yielding an expectation value of ($1.48 \pm 0.06$) counts per \Rn\ decay. The background count rate above $\unit{50}{keV}$ scatters around one count per day for the different proportional counters. Thus, a minimum detectable activity of \linebreak \mbox{$\sim\unit{20}{\upmu Bq}$} can be achieved.

Prior to a measurement, the emanated \Rn\ atoms had to be collected from the samples, concentrated and mixed with the counting gas. For this purpose, the samples were placed in gas-tight emanation vessels made of glass or stainless steel. We ensured that the pieces of a sample were not stacked in order to let radon escape freely from all surfaces. Ambient air was removed by pumping or flushing with radon-free carrier gases (in most cases helium). Then, the vessel was filled with the carrier gas and sealed and the emanated \Rn\ accumulated until the \Rn\ activity reached a sizable fraction of its equilibrium value. After typically one week, the mixture of carrier gas and \Rn\ atoms was pumped or flushed through a gas purifier to remove gaseous impurities. The \Rn\ atoms were trapped in an activated carbon column at liquid nitrogen temperature afterwards.

Larger samples, such as subsystems of the entire \linebreak XENON1T experiment, could not be placed in emanation vessels. In these cases, the carrier gas was usually filled directly into the gas-tight system. Such samples took the role of both, the emanation vessel and the investigated \Rn\ source, whereas the rest of the procedure remained \linebreak unchanged. Sometimes only a fraction of the filled carrier gas could be extracted due to limited pumping power. In such cases, the quoted activities were corrected for this reduced extraction efficiency, assuming that the radon was homogeneously distributed in the gas.

The concentrated \Rn\ sample was further processed in a sample purification system. The same system was also used to fill the proportional counters \cite{gas-line}. It featured several cold traps and a non-evaporable hot getter pump to remove unwanted trace-impurities that could impair the counter performance. In the final step, the gas sample was mixed with the counting gas and pushed into the counter by means a mercury column.

In some cases, measurements had to be done on samples that were previously exposed to xenon. The subsequent xenon out-gassing inhibited the use of miniaturized proportional counters due to their small volume and the difficulty to separate radon from xenon. In such cases we used electrostatic radon monitors with a significantly larger active volume \cite{radon_monitor, stefan_phd}. The positively charged \Rn\ daughters were collected on a silicon PIN diode biased with a negative high voltage with respect to the vessel's walls. All subsequent $\upalpha$-decays were recorded by the diode and the signal from $^{214}$Po was evaluated as it has the highest detection efficiency (approximately $\unit{30}{\%}$). We used two different monitors with a vessel size of $\unit{1}{liter}$ and $\unit{4}{liters}$, respectively. The background due to self-emanation of the monitor was negligible compared to the signal in all measurements. Even though the sensitivity of the radon monitors was about four times worse compared to the measurements with proportional counters, it was sufficient for our applications in XENON1T.

All results in this article are given with a combined uncertainty $\upsigma$ including statistical and systematical errors. If the result is compatible with zero within $1.645\, \upsigma$, a 90\,\% C.L. upper limit is given instead. Whenever a sample was additionally screened by $\upgamma$-ray spectroscopy, we quote the result obtained in \cite{gamma_screening_paper} and refer to the identifier used in that work as Radioassay-ID (RID).

\section{Material screening}
\label{individual_sample}
This section presents the results of samples which were measured during the preparation and construction phase of \linebreak XENON1T. We also list the supplier of the samples as it was not always possible to identify the manufacturer of the raw material.

\begin{table*}
\centering
\caption{\Rn\ emanation measurements of grade-1 titanium samples. The normalization to surface area and, where relevant, to weld seam length is also given. For comparison the \Ra\ bulk activity taken from \cite{gamma_screening_paper} is quoted and referred to as RID (Radioassay-ID) defined there.}
\begin{tabular}{cllllr@{ $\pm$ }lc}
\hline
ID & Sample & Supplier & Description & Treatment & \multicolumn{2}{c}{\Rn\ emanation rate} & \Ra\ activity \cite{gamma_screening_paper}\\
\hline
\#1 & Ti grade-1 & Supra & 14 plates & untreated & (120 & 30)\,$\upmu$Bq & RID \#40 \\
 & & Alloys  &  20.3\,cm$\times$20.3\,cm$\times $0.25\,cm & & (101 & 28)\,$\upmu$Bq/m$^2$ &  $<0.6$\,mBq/kg \\
 & & & Total: 1.2\,m$^2$ / 6.9\,kg  &  \multicolumn{2}{c}{}  &  \\ \hdashline
\#1a & & & same as sample \#1 & etched for 2 hours & (160 & 50)\,$\upmu$Bq &\\
 & & & & in 1.8\,\% HNO$_3$ & (140 & 40)\,$\upmu$Bq/m$^2$ &\\ \hline
\#2 & Ti grade-1 &  Nironit & 13 plates & untreated & (540 & 60)\,$\upmu$Bq & RID \#39\\
&  & & 20\,cm$\times$20\,cm$\times$0.2\,cm & & (510 & 5)\,$\upmu$Bq/m$^2$ & $<1.3$\,mBq/kg  \\
& & & Total: 1.1\,m$^2$ / 4.6\,kg & & (31 & 3)\,$\upmu$Bq/m &\\
& & &  17.4\,m of TIG-weld seam  &  \multicolumn{2}{c}{}  &  \\ \hdashline
\#2a & & &  7 plates of sample \#2 & etched for 2 hours  & (73 & 28)\,$\upmu$Bq & \\
& & & Total: 0.6\,m$^2$ / 2.4\,kg & in 0.6\,\% HF /  7.6\,\% HNO$_3$ & (130 & 50)\,$\upmu$Bq/m$^2$ &\\
& & & 9.4\,m of TIG-weld seam& ("pickling") & (7.8 & 2.9)\,$\upmu$Bq/m & \\ \hline
\#3 & Ti grade-1 & Nironit & 13 plates& untreated & (2980 & 200)\,$\upmu$Bq & RID \#38 \\
& & &  20\,cm$\times$20\,cm$\times$0.2\,cm  & & (2810 & 190)\,$\upmu$Bq/m$^2$ & ($1.1\pm0.4$)\,mBq/kg\\
& & & Total: 1.1\,m$^2$ / 4.6\,kg  &  \multicolumn{2}{c}{}  &  \\ \hdashline
\#3a & &  & 6 plates of sample \#3  & electropolished &  \multicolumn{2}{c}{$<39$\,$\upmu$Bq} & \\
& & &  Total: 0.5\,m$^2$ / 2.1\,kg & 30\,$\upmu$m surface removed & \multicolumn{2}{c}{$<80$\,$\upmu$Bq/m$^2$} & \\ \hline
\end{tabular}
\label{tab:titanium}
\end{table*}
\subsection{Metal samples}
\subsubsection{Titanium}
In an early phase of the project a cryostat made of grade-1 titanium was considered for XENON1T and the \Rn\ emanation rate of several titanium samples was measured. The results are given in Table \ref{tab:titanium}.

Sample \#1 was from Supra Alloys and samples \#2 and \#3 were from Nironit. There were 17.4 meters of TIG-wel\-ded\footnote{TIG: Tungsten Inert Gas} seam on the plates of sample \#2 to test the influence of welding on the \Rn\ emanation rate. We performed surface cleaning tests for all samples. Sample \#1a was treated for 2 hours in a $\unit{1.8}{\%}$ nitric acid (HNO$_3$) solution. Titanium itself is not soluble in nitric acid, but the acid may remove trace impurities, in particular $^{226}$Ra, from the surface. In contrast, sample \#2a was pickled for 2 hours in a 0.6\,\% hydrogen fluoride (HF) and 7.6\,\% nitric acid solution. This procedure dissolved titanium and consequently removed the upper few micrometers of the sample. In case of sample \#3a, 30\,$\upmu$m of the surface were removed by electro-polishing. All results are given as both, an absolute \Rn\ emanation rate and a rate normalized to the surface of the sample. For samples \#2 and \#2a we also normalize the result to the length of the weld seam. Note, however, that we could not distinguish whether \Rn\ came from the weld seam or from the surface. Thus, both normalizations cannot be true simultaneously and must be considered as upper limits.

Before any purification, the \Rn\ emanation rates of the grade-1 titanium samples exhibited large variations, which did not show up in the \Ra\ activity obtained from $\upgamma$-ray spectroscopy (last column in Table \ref{tab:titanium}), hinting at a surface contamination. The nitric acid treatment did not show any improvement (\#1a).  Pickling, in contrast, reduced the \Rn\ emanation rate. Under the assumption that the contamination was equally distributed among the 13 plates, we obtained a factor $4.0\pm1.6$ reduction by comparing the result of sample \#2 to sample \#2a. The most interesting sample was \#3 which showed a particularly high \Rn\ emanation rate. It disappeared completely after electro-polishing (\#3a). Evidently, a major part of the total \Ra\ activity of the sample was located on the surface. Thus, the true bulk activity must have been lower than reported in \cite{gamma_screening_paper} (RID \#38), since $\upgamma$-ray spectroscopy cannot resolve the spatial distribution of the radio-impurities in a sample and usually assumes that all activity is in the bulk. Sample \#3 and sample \#3a nicely illustrate the complementarity of \Rn\ assay technique and $\upgamma$-ray spectroscopy, and the importance to apply both methods.

The results of the titanium \Rn\ screening campaign suggested that electro-polishing can suppress the \Rn\ emanation rate of titanium to a negligible level. However, our titanium samples showed a too high uranium bulk contamination. Therefore, the XENON1T cryostat was made of stainless steel \cite{instrument_xenon}.

\subsubsection{Stainless steel}
\label{SS-measurements}

Stainless steel was mainly used for vessels and pipes in the XENON1T inner detector system. Parts of the TPC were also made of stainless steel, however, their surface area were small and the contribution to \Rn\ emanation was thus expected to be minor. Table \ref{tab:stainless} summarizes the results of all investigated stainless steel samples. Samples \#4 to \#9 were screened to investigate the contribution from stainless steel welds. Samples \#10 and \#11 were bellows which could be a potential \Rn\ source in stainless steel pipes. Samples \#12 to \#14 were related to heat exchangers and sample \#15 was the packing material of the xenon distillation column \cite{kr_distillation}. As in the case of titanium, we could not distinguish whether the measured \Rn\ emanation rate of TIG-welded samples originated from the surface or from the weld. Thus, in Table \ref{tab:stainless} we give both normalizations in addition to the absolute \Rn\ emanation rate.
\begin{table*}
\centering
\caption{\Rn\ emanation results of various stainless steel (SS) samples. Where relevant, the normalization to surface area and to weld seam length is also given.}
\begin{tabular}{cllllr@{ $\pm$ }l}
\hline
ID & Sample & Supplier & Description & Treatment & \multicolumn{2}{c}{\Rn\ emanation rate}\\
\hline
\#4 & Plates with welds& CRIOTEC Impianti S.p.A. & 5 plates w. 8.25\,m TIG-weld seam & untreated & (190 & 30)\,$\upmu$Bq\\
  &  (304 or 316L) & & 17.9\,cm x 17.6\,cm x 1\,cm & & (560 & 100)\,$\upmu$Bq/m$^2$\\
  & & & Total: 0.33\,m$^2$ / 12.9\,kg & & (23 & 4)\,$\upmu$Bq/m \\ \hline
\#5 & Tubes with welds & ALCA Technology S.r.l. & 3 tubes w. 4.8\,m TIG-weld seam  & untreated & (214 & 26)\,$\upmu$Bq\\
  & (316L) & & Outer diameter: 10.18\,cm& & (380 & 50)\,$\upmu$Bq/m$^2$\\
  & & & Thickness of wall: 0.34\,cm  & & (45 & 5)\,$\upmu$Bq/m\\
  & & & Length: 30\,cm & & \multicolumn{2}{c}{}\\
  & & & Total: 0.56\,m$^2$ / 7.2\,kg & & \multicolumn{2}{c}{}\\ \hdashline
\#5a & Tubes with welds & ALCA Technology S.r.l. & same as sample \#5 & electro-polished & (47 & 19)\,$\upmu$Bq\\
  & (316L) & & & & (80 & 30)\,$\upmu$Bq/m$^2$\\
  & & & & & (10 & 4)\,$\upmu$Bq/m\\ \hline
\#6 & Tubes & ALCA Technology S.r.l. & 3 tubes & untreated & (160 & 50)\,$\upmu$Bq\\
  & (316L) & & Outer diameter: 10.18\,cm& & (290 & 90)\,$\upmu$Bq/m$^2$\\
  & & & Thickness of wall: 0.34\,cm  & & \multicolumn{2}{c}{}\\
  & & & Length: 30\,cm & & \multicolumn{2}{c}{} \\
  & & & Total: 0.56\,m$^2$ / 7.2\,kg & & \multicolumn{2}{c}{} \\ \hline
\#7 & Tube with welds & Lamm's machine Inc. & Tube with 3.6\,m TIG-weld seam & untreated & \multicolumn{2}{c}{$\leq$38\,$\upmu$Bq} \\ 
  & (304L) & & Outer diameter: 10.16\,cm& & \multicolumn{2}{c}{$\leq$200\,$\upmu$Bq/m$^2$}\\
  & & & Thickness of wall: 0.2\,cm  & & \multicolumn{2}{c}{$\leq$11\,$\upmu$Bq/m}\\
  & & & Length: 34\,cm & & \multicolumn{2}{c}{} \\
  & & & Total: 0.19\,m$^2$ / 1.5\,kg & & \multicolumn{2}{c}{} \\ \hline
\#8 & Tube with welds & Lamm's machine Inc. & Tube with 3.6\,m TIG-weld seam & electro-polished & (52 & 23)\,$\upmu$Bq\\ 
  & (304L) & & Outer diameter: 10.16\,cm& & (270 & 120)\,$\upmu$Bq/m$^2$\\
  & & & Thickness of wall: 0.2\,cm  & & (14 & 6)\,$\upmu$Bq/m\\
  & & & Length: 34\,cm & & \multicolumn{2}{c}{} \\
  & & & Total: 0.19\,m$^2$ / 1.5\,kg & & \multicolumn{2}{c}{} \\ \hline
\#9 & Tube with welds & Lamm's machine Inc. & Tube with 3.6\,m TIG-weld seam & etched & (57 & 17)\,$\upmu$Bq \\ 
  & (304L) & & Outer diameter: 10.16\,cm & and & (300 & 90)\,$\upmu$Bq/m$^2$ \\
  & & & Thickness of wall: 0.2\,cm  & electro-polished & (16 & 4)\,$\upmu$Bq/m\\
  & & & Length: 34\,cm & & \multicolumn{2}{c}{} \\
  & & & Total: 0.19\,m$^2$ / 1.5\,kg & & \multicolumn{2}{c}{} \\ \hline
\#10 & Stainless steel bellow & Streas S.r.l. & Bellow with one CF40 flange & untreated & (130 & 40)\,$\upmu$Bq \\
  & & & 1m long, inner diameter 35\,mm & \multicolumn{2}{c}{} \\ \hline
\#11 & Stainless steel bellow & Streas S.r.l. & Bellow with one CF100 flange& electro-polished & \multicolumn{2}{c}{$\leq$160\,$\upmu$Bq}\\ 
  & & & 0.4m long, inner diameter 100\,mm & \multicolumn{2}{c}{} \\ \hline
\#12 & Large heat exchanger & GEA Group & Type FG5X12-60: & untreated & (510 & 50)\,$\upmu$Bq\\
  &  &  & 60 SS plates (338\,mm $\times$ 130\,mm) &  &  \multicolumn{2}{c}{} \\
  & & & brazed with copper alloy & &  \multicolumn{2}{c}{} \\ \hline
\#13 & Small heat exchanger & GEA Group & Type FG3X8-20: & etched for & (134 & 24)\,$\upmu$Bq \\
 & combined with & & 20 SS plates (226\,mm $\times$ 86\,mm) & $\sim12$h with & \multicolumn{2}{c}{}\\
  &   large heat exchanger & &  brazed with copper alloy  & 1.8\,\% HNO$_3$ & \multicolumn{2}{c}{} \\ 
    &  & & plus type FG5X12-60 (see above) & &  \multicolumn{2}{c}{} \\ \hline
\#14 &  High-purity electrical & D.A.T.E. (Developpement & custom-designed heater complying & etched for & (70 & 27)\,$\upmu$Bq\\ 
& heater with large & et Applications des & with very high purity standards & $\sim15$\,min with & \multicolumn{2}{c}{}\\
& heat transfer surface & Techniques de L'Energie) & & 2\,\% HNO$_3$ & \multicolumn{2}{c}{}\\
\hline
\#15 & Stainless steel & Sulzer Ltd. &  55 structured packings, type EX & untreated & (48 & 20)\,$\upmu$Bq \\ 
 & packing material & & 
 0.095\,m$^2$/piece, total surface: 5.2\,m$^2$& & (9 & 4)\,$\upmu$Bq/m$^2$ \\ \hline
\end{tabular}
\label{tab:stainless}
\end{table*}

Sample \#4 and sample \#5 had a similar \Rn\ emanation rate, although the weld seam length of sample \#4 was significantly longer. This pointed to a subdominant contribution of the welds. That result was further supported by a test of sample \#6. It had no welds at all, but showed a similar \Rn\ emanation rate as the welded sample \#5 (normalized to its surface). Subsequent electro-polishing of sample \#5 reduced the measured \Rn\ emanation rate by a factor $3.4\pm1.2$ (\#5a).

We further investigated stainless steel samples TIG-wel\-ded by Lamm's Machine Inc., the company that built the cryogenic system for XENON1T. We did not test any unwelded samples, but three welded samples with different surface treatments (\#7 -- \#9). The results normalized to the surface area were in agreement with samples \#5 and \#6 from ALCA Technology S.r.l.  Note that an upper limit was found for the untreated sample (\#7). Hence, a further \Rn\ reduction by cleaning attempts was not measurable within our sensitivity. Again it was confirmed that stainless steel TIG-welds do not represent a notable additional source of \Rn, which is in tension with findings from other experiments \cite{SimgenBX, GERDA_welds}.

The measurements of the bellows were motivated by the relatively high \Rn\ emanation rate of the cryogenic pipe (\#52 in Table \ref{tab:subsystems}). This was a six-fold stainless steel pipe-in-pipe system that connected the XENON1T cryostat to the cryogenic system (see section \ref{sub:integration_subsystems} and \cite{instrument_xenon}). It contained about 10 meters of stainless steel bellows in the pipes to compensate for thermal shrinkage. We tested two spare bellows from the same supplier with $\unit{35}{mm}$ and $\unit{100}{mm}$ inner diameter, respectively (\#10 and \#11). Although we found a small positive signal for sample \#10, the result indicated that the bellows did not constitute the main \Rn\ source of the cryogenic pipe.

We also tested two heat exchangers, which were used to evaporate and re-condense xenon in the purification loop. They were made of stainless steel plates brazed with a copper alloy. We measured the larger one (\#12) prior to any purification and found a \Rn\ emanation rate of $\unit{(510 \pm 50)}{\upmu Bq}$. Subsequently, we cleaned it by exposing all internal surfaces to a $\unit{1.8}{\%}$ nitric acid solution for about 12 hours. The same treatment was done for a smaller heat exchanger of the same type, which, however, was not measured before. After the treatment both heat exchangers were measured together (\#13) and yielded a result of $(134 \pm 24)$\,$\upmu$Bq. Clearly, the treatment was effective despite the rather weak concentration of nitric acid.  The heat exchangers were combined with a high purity electrical heater (\#14). Its \Rn\ emanation rate was found to be $\unit{(70 \pm 27)}{\upmu Bq}$.

The last sample (\#15) in Table \ref{tab:stainless} was from the XENON1T xenon distillation column for krypton removal \cite{kr_distillation}. The column was filled with structured stainless steel packings to increase the contact surface between the gaseous and liquid phase of xenon. Since the same column was used for radon removal tests \cite{xenon_rn_dist} (see section \ref{sub:reduction}), we were interested in its \Rn\ emanation rate. Indeed, all 55 packings together only emanated $\unit{(48\pm20)}{\upmu Bq}$, which was an excellent result for the rather large sample surface (5.2~m$^2$).

\subsection{Gas purification system}
\label{XEPUR}
\subsubsection{Gas purifiers}
\label{subsec_getters}

\begin{table*}[ht]
\centering 
\caption{\Rn\ emanation rates of four noble gas purifiers from the company SAES. PS4-MT50-R535 is identical to PS4-MT50-R2, but received a new commercial label.}  
\begin{tabular}{llcccc}
\hline 
ID & Model & Mass of active material & Cold state activity [mBq] & Hot state activity [mBq] & Used in XENON1T \\ 
\hline 
\#16 & MonoTorr PS4-MT50-R2 & $\sim4$\,kg & $0.61 \pm0.04$ & $1.17 \pm0.15$ & yes\\ 
\#17 & MonoTorr PS4-MT50-R535 & $\sim4$\,kg & -- & $0.24 \pm 0.03$ & yes\\ 
\#18 & MonoTorr PS4-MT3-R2 & $\sim0.5$\,kg & $0.09 \pm 0.03$ & $0.09 \pm 0.03$ & no\\ 
\#19 & MonoTorr PS3-MT3-R2 & $\sim0.5$\,kg & -- & $\leq$0.015 & no\\  
\hline
\end{tabular}      
\label{tab:getter}
\end{table*}

In order to maintain its ultra-high chemical purity, the xenon in XENON1T was continuously cleaned by SAES gas purifiers. Two of them were used in parallel to provide the required purification efficiency. Each purifier contained a porous, highly chemically-active zir\-co\-nium-alloy in two cartridges. The larger was operated at 400~$^\circ$C, while the smaller one was kept at room temperature and acted as a dedicated hydrogen removal unit. Altogether, $\sim4$\,kg of active material was used in each purifier. 

We measured the two gas purifiers of XENON1T (\#16 and \#17) and two smaller models (\#18 and \#19 with about eight times smaller active mass) which used the same alloy, but which had no hydrogen removal unit. The results are presented in Table \ref{tab:getter}. We measured the \Rn\ emanation rate in two different thermal conditions; at room temperature (cold state) and at operating temperature (hot state). While only the hot state was relevant for the experiment, the measured rate of the cold state could have given insight into the \Rn\ emanation process relevant for these porous materials. An enhanced \Rn\ release rate at elevated temperature would have given evidence for diffusion-driven emanation. The two gas purifiers used in XENON1T differed significantly in their hot state emanation rate (\#16 emanated almost five times more \Rn\ than \#17). The two smaller gas purifiers showed a significantly lower \Rn\ emanation rate. For sample \#16, the cold state emanation was reduced by a factor $1.91\pm0.27$ in comparison with its hot state. In contrast, for sample \#18 no difference between the hot and cold state was observed within the measurement uncertainty.

The \Rn\ emanation rate of the gas purifiers is not fully understood, but the large difference for identical models suggested that it depends on the purity of the raw materials. That opens up the possibility for further \Rn\ reduction by material screening.

\subsubsection{Recirculation pumps}
\label{recirculation-pumps}

XENON1T used customized QDrive piston pumps from \linebreak Chart Industries for xenon gas recirculation \cite{qdrive-pump}. Three \linebreak pumps (\#20, \#21 and \#22) were measured for their \Rn\ emanation rate. The results are summarized in Table \ref{tab:QDrive1}. After a mechanical failure of QDrive pump C204, it was sent back to the manufacturer for repair. Afterwards, its \Rn\ emanation rate was lower by more than a factor two (\#20a), probably due to the replacement of a dirty polyester resin (see text below and Table \ref{tab:QDrive3}).\\

\begin{table*}
\centering
\caption{\Rn\ emanation measurements of xenon recirculation pumps.} 
\begin{tabular}{lllr@{ $\pm$ }lcc}
\hline
ID & Sample & Description & \multicolumn{2}{c}{\Rn\ emanation} & Used & Comment \\
  & & & \multicolumn{2}{c}{rate [mBq]} & in XENON1T & \\
\hline
\#20  & QDrive pump 2S132CX & Serial No. C204 & 5.2 & 0.2 & no &\\
\#20a & QDrive pump 2S132CX & Serial No. C204 after repair & 2.5 & 0.1 & yes &\\ 
\#21  & QDrive pump 2S132CX & Serial No. C205 & 3.5 & 0.2 & no &\\
\#22  & QDrive pump 2S132CX & Serial No. C206 & \hspace{5ex} 4.5 & 0.2 & yes &\\
\hline
\#23  & \multicolumn{2}{l}{Magnetically coupled piston pump \cite{muenster-pump}}  & 0.29 & 0.09 & yes & replacing all QDrive pumps \\
\hline
\end{tabular}
\label{tab:QDrive1}
\end{table*}

\begin{table*}
\centering
\caption{\Rn\ emanation measurements of all parts of a QDrive recirculation pump prior to assembly.}
\begin{tabular}{lllr@{ $\pm$ }l}
\hline
ID & Sample & Description & \multicolumn{2}{c}{\Rn\ emanation rate} \\
\hline
\#24 & 2 stators & silicon steel frame, Cu wire winding, polyester impregnate & \hspace{4ex} (2.99 & 0.15)\,mBq \\
\#25 & 2 pistons & made of brass & (0.68 & 0.06)\,mBq \\
\#26 & 2 magnet cores & each equipped with 8 magnets & (0.28 & 0.11)\,mBq \\
\#27 & 4 flexure assemblies & silicon steel, stainless steel & (0.132 & 0.021)\,mBq \\
\#28 & Polyester lacing & to fix Cu wire winding & \multicolumn{2}{c}{$<0.053$\,mBq}\\
\#29 & Remaining small items & 3 other types of Epoxy, Screws, Nuts, Spacers & \multicolumn{2}{c}{$<0.020$\,mBq}\\ \hline
\end{tabular}
\label{tab:QDrive2}
\end{table*}

\begin{table*}
\centering
\caption{Results of \Rn\ emanation measurement to identify and replace the dominant \Rn\ sources in the QDrive pump.}
\begin{tabular}{lllr@{ $\pm$ }l}
\hline

\#30 & Bare magnets & 6 pieces: 1.55\,cm x 1.88\,cm x 1.88\,cm each & \multicolumn{2}{c}{$<0.021$\,mBq}\\
& &  Total surface: 112\,cm$^2$ & \multicolumn{2}{c}{$<1.9$\,mBq/m$^2$}\\ \hdashline
\#31 & Magnets with epoxy coating & 8 pieces: 1.55\,cm $\times$ 1.88\,cm $\times$ 1.88\,cm each&  (0.37 & 0.04)\,mBq \\ 
& &  Total surface: 150\,cm$^2$ & (24.8 & 2.9)\,mBq/m$^2$ \\ \hdashline
\#32 & Alternative epoxy coating for magnets & cured on Cu substrate & (0.093 & 0.017)\,mBq\\
 & (from Magnet Coating Engineering) & Surface: 618\,cm$^2$ & (1.51 & 0.28)\,mBq/m$^2$\\ \hline
\#33 & 1.1mm copper wire for magnet winding & 1.68\,kg & \multicolumn{2}{c}{$<0.097$\,mBq}\\  \hline
\#34 & Wingard \& Co. silicon steel & 6 plates (9.3\,cm $\times$ 8.9\,cm $\times$ 0.05\,cm)  & (0.240 & 0.040)\,mBq\\ 
& & fulfills ASTM A677 standard & (14.1 & 2.2)\,mBq/m$^2$\\
& & Total: 167\,cm$^2$ / 29.4\,g & \multicolumn{2}{c}{}\\ \hline
\#35 & Dolphon CC-1105 HTC & solventless polyester resin& (0.70 & 0.05)\,mBq\\
& & 396~cm$^2$ / 95.4\,g & (17.7 & 1.2)\,mBq/m$^2$ \\ \hdashline
\#36 & Elantas GRC 59-25 & low viscosity hermetic varnish & (0.027 & 0.012)\,mBq\\
& (alternative to Dolphon CC-1105 HTC) & 398\,cm$^2$ / 51.9\,g & (0.76 & 0.30)\,mBq/m$^2$ \\\hline
\end{tabular}
\label{tab:QDrive3}
\end{table*}

Since the contribution of the recirculation pumps presented a major fraction of XENON1T's total \Rn\ budget (see section \ref{sub:rn_budget}), we performed further investigations to understand the origin of the observed \Rn\ emanation rate. For this purpose, we screened most of the individual components of a yet unassembled QDrive pump (see Table \ref{tab:QDrive2}). We found that the stators of the pump's electrical motor (\#24) were the dominant \Rn\ sources, followed by the pistons (\#25) and the magnet cores (\#26). 

To study the origin of the identified \Rn\ sources, we further investigated their constituent parts (see Table \ref{tab:QDrive3}). First, we noticed that the bare magnets did not emanate a lot of \Rn\ (\#30). The \Rn\ emanation rate rather originated from the epoxy coating as can be seen from the comparison to sample \#31. Note, however, that the dirty coating of sample \#31 was only used for prototype pumps. In the pumps of XENON1T it was already replaced by an alternative much cleaner coating (\#32), which we identified by our screening effort. The stator was a ring-shaped structure of silicon steel with four electromagnets, each formed by a copper coil which was held in place by polyester resin. The wire used for the copper coil showed no measurable \Rn\ emanation rate (\#33), whereas a significant emanation rate was found for the silicon steel (\#34) and the polyester resin (\#35). The latter one was responsible for a large fraction of the stator's \Rn\ emanation rate. Later, we identified a cleaner alternative (\#36). The new resin was applied in the repaired C204 pump. It may explain the observed \Rn\ reduction (\#20 and \#20a), if one considers that the surface to volume ratio of the resin inside the pump was larger than for the aliquots of sample \#35 and \#36. Moreover, the used amount of resin in a pump was hard to quantify and fluctuated among different pumps.

Motivated by the relatively large \Rn\ emanation rate of the QDrive pumps, we followed an approach of EXO-200 \cite{exo200-pump} to build a cleaner magnetically coupled piston pump. The new pump was developed within the XENON collaboration together with groups from the nEXO collaboration \cite{muenster-pump}. Its $^{222}$Rn emanation rate was found to be 
($0.29 \pm 0.09$) mBq (\#23 in Table \ref{tab:QDrive1}), an order of magnitude lower than the results obtained for the QDrive pumps. The new pump was successfully installed for the last data taking phase of XENON1T and its impact on the \Rn\ budget is discussed in section \ref{sub:reduction}.

\subsection{Other samples}
\label{components}

This section presents \Rn\ emanation measurements of TPC components and other samples measured for the XENON1T experiment. The complexity of the XENON1T TPC made it impossible to screen every component. Therefore, we focused on samples which either cover a large surface area inside the TPC or are known to be potential \Rn\ sources.  We investigated the light sensors (\#37 and \#38), their cables and connectors (\#39 to \#43), the potting material for the cable feedthroughs (\#44), the PTFE reflectors of the TPC (\#45), copper of the field shaping rings (\#46) as well as a $^{220}$Rn calibration source (\#47). A detailed description of their usage in XENON1T can be found in \cite{instrument_xenon}. The results are summarized in Table \ref{tab:other_samples}.

XENON1T used Hamamatsu R11410-21 photo\-mul\-ti\-pli\-er tubes (PMTs) \cite{PMT-paper} as light sensors (\#37). Their \Rn\ emanation rate was measured in a helium-free environment and we used neon as carrier gas. Helium would have penetrated into the PMTs, creating an unacceptably high rate of afterpulses \cite{hamamatsu}. We also measured the \Rn\ emanation rate of the PMT high voltage divider circuits (base) used to read out the signal and to supply the high voltage (\#38). They consisted of a printed circuit board made of Cirlex, and several resistors and capacitors soldered onto it. It should be noted that after the measurement a different type of resistors was selected for the boards used in XENON1T. It proved to have a comparable \Ra\ activity, but no dedicated \Rn\ emanation test was performed. With the results given in Table \ref{tab:other_samples} and taking into account that XENON1T used 248 PMTs, we estimated a total contribution of $\unit{(1.08\pm0.26)}{mBq}$ from the PMTs and their bases. Note that the \Ra\ activity in the PMTs (RID \#69 in \cite{gamma_screening_paper}) was much higher, which means that less than a percent was emanated.

\begin{table*}
\centering
\caption{\Rn\ emanation measurements of various other samples. For comparison the \Ra\ bulk activity taken from \cite{gamma_screening_paper} is quoted and referred to as RID (Radioassay-ID) defined there.}
\begin{tabular}{llllr@{ $\pm$ }lr@{ $\pm$ }l}
\hline
ID & Sample & Supplier & Description & \multicolumn{2}{c}{\Rn\ emanation rate} & \multicolumn{2}{c}{\Ra\ activity \cite{gamma_screening_paper}} \\ \hline
\#37 & R11410 PMTs & Hamamatsu & 29 pieces, mixture of & (58 & 28)\,$\upmu$Bq & \multicolumn{2}{c}{RID \#69} \\
& & & low performance PMTs & (2.0 & 1.0)\,$\upmu$Bq/PMT & (600 & 100)\,$\upmu$Bq/PMT\\
& & & and mechanical samples & \multicolumn{2}{c}{} &   \multicolumn{2}{c}{} \\  \hdashline
\#38 & PMT bases & Fralock & 55 pieces, made of Cirlex  & (129 & 25)\,$\upmu$Bq & \multicolumn{2}{c}{RID \#94}\\
& & & with soldered resistors & (2.4 & 0.5)\,$\upmu$Bq/piece & (15 & 2)\,$\upmu$Bq/piece \\ 
& & &  and capacitors &  \multicolumn{2}{c}{} & \multicolumn{2}{c}{}  \\  \hline
\#39 & Kapton & Accu-Glass & 30 AWG solid core wire &  \multicolumn{2}{c}{$\leq$35\,$\upmu$Bq} & \multicolumn{2}{c}{RID \#57}\\
&  single wire cable & & 100\,m / 0.076\,kg           & \multicolumn{2}{c}{$\leq$0.35\,$\upmu$Bq/m} &  (4000 & 1000)\,$\upmu$Bq/kg \\
& & & for high voltage supply  & \multicolumn{2}{c}{$\leq$460\,k$\upmu$Bq/kg} & \multicolumn{2}{c}{}\\
& & & (used in XENON1T) & \multicolumn{2}{c}{} & \multicolumn{2}{c}{} \\ \hdashline
\#40 & Kapton coaxial cable & Accu-Glass & 30 AWG, 50 $\Omega$ cable  &  \multicolumn{2}{c}{$\leq$25\,$\upmu$Bq} & \multicolumn{2}{c}{} \\
& & & 100\,m / 0.55\,kg &  \multicolumn{2}{c}{$\leq$0.25\,$\upmu$Bq/m} & \multicolumn{2}{c}{}\\
& & & for signal readout & \multicolumn{2}{c}{$\leq$45\,$\upmu$Bq/kg} & \multicolumn{2}{c}{} \\
& & & (not used in XENON1T) & \multicolumn{2}{c}{} & \multicolumn{2}{c}{} \\ \hdashline
\#41 & PTFE coaxial cable & Huber \& Suhner & RG196  & \multicolumn{2}{c}{$\leq$44\,$\upmu$Bq}  & \multicolumn{2}{c}{RID \#59}\\
& & & 78\,m / 0.594\,kg & \multicolumn{2}{c}{$\leq$0.56\,$\upmu$Bq/m} & (1000 & 300)\,$\upmu$Bq/kg\\
& & & for signal readout &  \multicolumn{2}{c}{$\leq$74\,$\upmu$Bq/kg} & \multicolumn{2}{c}{}\\
& & & (not used in XENON1T) & \multicolumn{2}{c}{} & \multicolumn{2}{c}{} \\ \hdashline
\#42 & PTFE coaxial cable & koax24 & RG196 & \multicolumn{2}{c}{$\leq$58\,$\upmu$Bq}  & \multicolumn{2}{c}{RID \#55 and \#56} \\
& & & 182\,m / 1.59\,kg & \multicolumn{2}{c}{$\leq$0.32\,$\upmu$Bq/m}  &  (400 & 200)\,$\upmu$Bq/kg\\
& & & for signal readout  & \multicolumn{2}{c}{$\leq$36\,$\upmu$Bq/kg} &\multicolumn{2}{c}{} \\
& & & (used in XENON1T) & \multicolumn{2}{c}{} & \multicolumn{2}{c}{} \\ \hline
\#43 & D-sub type & Accu-Glass & 1200 male and female pieces & \multicolumn{2}{c}{$\leq$47\,$\upmu$Bq} & \multicolumn{2}{c}{} \\
& contact pins  & &  made of Cu/Be and Cu/bronze & \multicolumn{2}{c}{$\leq$0.039\,$\upmu$Bq/piece} & \multicolumn{2}{c}{}\\ \hline
\#44 & Epoxy for potting & Reliable Hermetic & 3 discs & \multicolumn{2}{c}{$\leq$51\,$\upmu$Bq}  & \multicolumn{2}{c}{} \\
& & Seals & 9.5\,mm thick, 95\,mm diameter& \multicolumn{2}{c}{$\leq$1000\,$\upmu$Bq/m$^2$} & \multicolumn{2}{c}{} \\
& & & 506\,cm$^2$ / 191\,cm$^3$ / 0.47\,kg  & \multicolumn{2}{c}{} & \multicolumn{2}{c}{}\\ \hline
\#45 & PTFE panels & Amsler \& Frey & 67 pieces, total: 4.06\,m$^2$ / 31.9\,kg & (97 & 21)\,$\upmu$Bq &  \multicolumn{2}{c}{RID \#50}\\
& & & length: 19\,cm -- 24.5\,cm & (24 & 5)\,$\upmu$Bq/m$^2$ &  \multicolumn{2}{c}{$<120\,\upmu$Bq/kg}\\
& & & width: 13\,cm -- 19\,cm & (3.0 & 0.7)\,$\upmu$Bq/kg  & \multicolumn{2}{c}{}\\
& & & thickness: 0.5\,cm -- 1.6\,cm & \multicolumn{2}{c}{} & \multicolumn{2}{c}{}  \\ \hline
\#46 & Copper rods  & Luvata & 150 pieces, each 15\,cm long & \multicolumn{2}{c}{$\leq$25\,$\upmu$Bq} & \multicolumn{2}{c}{} \\
& & & and 2\,cm diameter & \multicolumn{2}{c}{$\leq$17\,$\upmu$Bq/m$^2$} & \multicolumn{2}{c}{} \\
& & & 1.5\,m$^2$ / 7069\,cm$^3$ / 56.7\,kg & \multicolumn{2}{c}{$\leq$0.44\,$\upmu$Bq/kg} & \multicolumn{2}{c}{} \\ \hline
\#47 & $^{220}$Rn calibration & PTB & electro-deposited $^{228}$Th &  \multicolumn{2}{c}{$\leq$46\,$\upmu$Bq } & \multicolumn{2}{c}{} \\
& source \cite{220Rn-source} \footnotemark & & on stainless steel disc &\multicolumn{2}{c}{}  & \multicolumn{2}{c}{} \\ \hline

\end{tabular}
\label{tab:other_samples}
\end{table*}

We also investigated the cables, which were used inside the detector. We measured the Kapton-insulated high voltage cable (\#39) and three types of coaxial cable, one with Kapton insulation (\#40) and two with PTFE insulation (\#41 and \#42). The \Rn\ emanation rate for all cables was below the detection limit. The \Ra\ activity of the cables, \cite{gamma_screening_paper} normalized to their mass was at least ten times larger than its \Rn\ emanation rates for each cable sample. This indicated that the \Rn\ sources were located in the inner part of the cables. The impact of the cables on the \Rn\ budget is discussed in more detail in section \ref{sub:integration_subsystems}.  Different high voltage cable sections were connected at the top of the TPC and just in front of the vacuum feedthroughs with D-sub type pins which showed no measurable \Rn\ emanation rate (\#43). The cables were fed to the air-side via potted feedthroughs. The epoxy used for this potting (\#44) was measured. For about half a kilogram of material, we found an upper limit of $\leq51$~$\upmu$Bq.

The reflective walls of the XENON1T TPC were formed by diamond-shaved PTFE panels. We measured (non \linebreak diamond-shaved) leftover pieces from the fabrication of these panels, corresponding to a surface area of $\geq$50\,\% of the PTFE surface in XENON1T.
The results (see \#45) indicated that PTFE was a sub-dominant \Rn\ source inside the TPC.

We also measured copper rods from the same batch as the TPC field-shaping rings (\#46), for which we found no detectable signal. Finally, we measured the \Rn\ release rate of a $^{220}$Rn source that was used as a calibration source for XENON1T (\#48) \cite{220Rn-source}. If the source had also released some \Rn, it would have required a long waiting time after each usage until the \Rn\ had decayed. Our measurement showed that its \Rn\ emanation rate was below the detection limit\footnotemark[\value{footnote}] and negligible with respect to the other \Rn\ sources.
 
\footnotetext{The limit presented here differs slightly from the result published in  \cite{220Rn-source} ($\leq 55~\upmu$Bq) due to a data re-evaluation.}

\section{XENON1T results}
\label{sub_system}

\subsection{Measurements of subsystems}
\label{sub:integration_subsystems}

\begin{figure*}[ht!]
\centering
\includegraphics[height=7cm,]{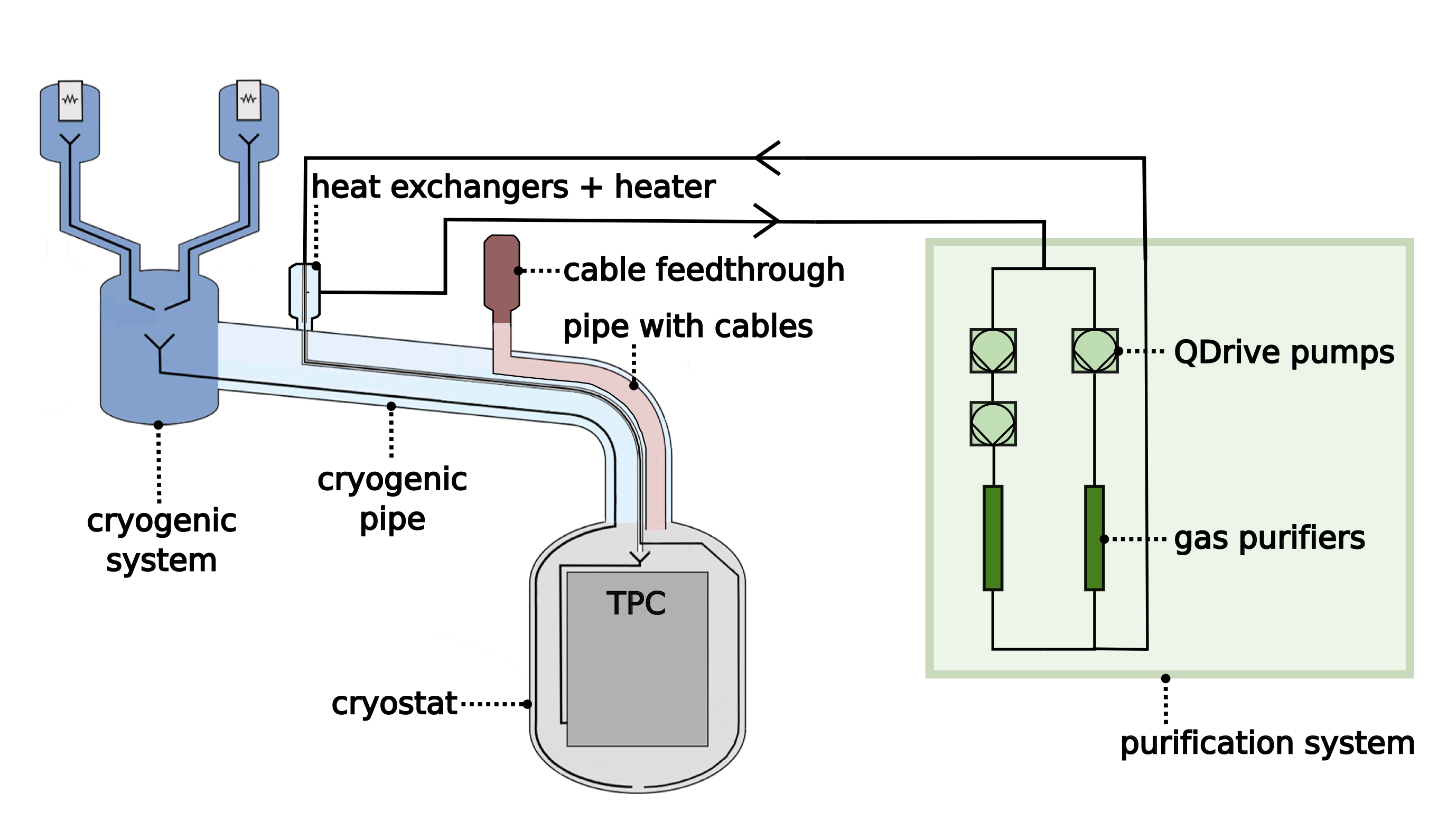}
\caption{Schematic setup of XENON1T (not to scale). The colors indicate different sub-components and are also used in Figure \ref{fig:piechart}, which shows their individual contributions to the overall $^{222}$Rn rate.}

 \label{fig:xenon_system}
\end{figure*}

\begin{figure*}[ht!]
\centering
\includegraphics[height=7cm]{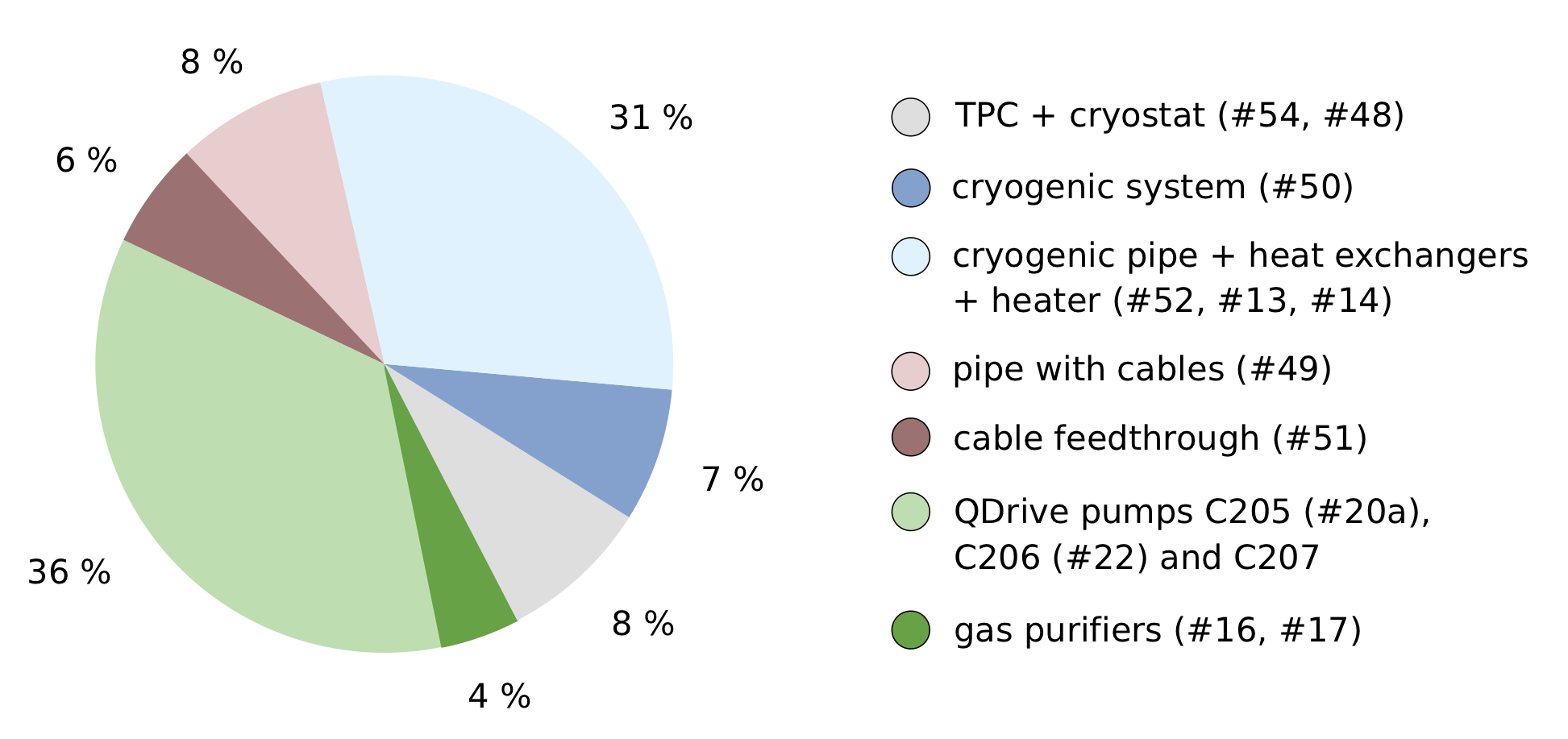}
\caption{The different sub-system contributions to the overall $^{222}$Rn emanation rate in XENON1T. The colors correspond to those used in Figure \ref{fig:xenon_system}. The numbers in the brackets refer to the item numbers. QDrive pump C207 was not measured. Its \Rn\ emanation rate was estimated (see text).}

 \label{fig:piechart}
\end{figure*}

In the following, we briefly describe the XENON1T inner detector systems and report on the \Rn\ emanation results of either individual subsystems or of combined measurements among them. More details on the measurements themselves can be found in \cite{stefan_phd, natascha_master}. We only assayed those subsystems that were continuously in contact with xenon during the data acquisition periods and, therefore, contributed to the \Rn\ budget. A schematic view of the subsystems most relevant for \Rn\ emanation is shown in Figure \ref{fig:xenon_system}.

The TPC was hosted in a double-walled, vacuum-in\-su\-la\-ted stainless steel cryostat, which was closed at the top by a dome. The dome was in turn connected to the cryogenic pipe, which also contained the cables and guided them to the electric feedthrough.  Evaporated xenon which reached the cryogenic system got liquefied and was returned back to the cryostat. In a second loop, the xenon gas passed through the purification system that contained the recirculation pumps and the gas purifiers. Note that a major part of the XENON1T infrastructure will be re-used for its upgrade XENONnT, so the obtained results will be relevant for the future as well \cite{XENONnT_sensitivity}.

\begin{table*}[ht]
\centering
\caption{Results from measurements of several subsystems of the XENON1T setup. The cryogenic pipe and the TPC were not measured directly. Their \Rn\ emanation rates were inferred indirectly by subtracting the results from two measurements, respectively.}
\begin{tabular}{llcc}
\hline 
ID & component & activity [mBq] & comment \\ 
\hline 
\#48 & cryostat (inner vessel) & $1.8 \pm 0.3$ &  \\ 
\#49 & pipe with cables & $2.7 \pm0.2$ & \\
\#50 & cryogenic system & $ 2.4  \pm  0.3$ & \\  
\#51 & cable feedthroughs & $ 1.9\pm 0.2$ &\\ 
\#52 & cryogenic pipe (without cable pipe and cables) & $ 9.4 \pm  1.0$ & indirect measurement\\
\#13 and \#14 & heat exchangers and heater & $0.20 \pm 0.04$ & from Table \ref{tab:stainless} \\ \hdashline
     & inner detector volume without TPC & $18.4 \pm 1.0$ & \\ 
     & (sum of all items above) &  &\\ \hline
\#53 & inner detector volume with TPC & $ 19.3 \pm  2.1$ &\\ \hline
\#54 & TPC & $\leq4.4$ & indirect measurement: difference between\\
     &     & & inner detector volume with and without TPC\\ 
\hline
\end{tabular} 
\label{tab:subsystems}
\end{table*}

The results of the measurements are listed in Table \ref{tab:subsystems}. After electro-polishing its inner surface, the cryostat (\#48) was measured at the fabrication site (ALCA Technology S.r.l.).  From the result of sample \#5a (see Table \ref{tab:stainless}) we predicted a \Rn\ emanation rate of $\unit{(80\pm30)}{\upmu Bq/m^2}$ for electro-polished stainless steel under the assumption of no contribution from the weld seam. Thus, we expected $\unit{(610\pm230)}{\upmu Bq}$ for the $\unit{7.6}{m^{2}}$ surface of the cryostat. This was about one third of what we have measured (\#48). The discrepancy may come from the fact that a large vessel cannot be cleaned as easily as small-size samples.

The cable pipe contained not only the cables for all 248 XENON1T channels, but also about 200 extra channels foreseen for the upgrade to XENONnT. Altogether, there were $\unit{4.1}{km}$ of PTFE insulated coaxial PMT signal cables and $\unit{4.5}{km}$ of Kapton insulated high voltage cables in the detector. From the results of sample \#39 and sample \#42 presented in section \ref{components}, we derived an upper limit of \mbox{$\leq 2.9$}\,mBq for the cables alone. For the entire cable pipe we measured a positive number of $(2.7\pm0.2)$\,mBq (\#49), which is compatible with that limit. Similar contributions to the overall \Rn\ budget were found to originate from the cryogenic system (\#50) and from the cable feedthroughs (\#51), respectively.  

The cryogenic pipe (\#52) consisted of a $\unit{250}{mm}$ diameter pipe enclosed in a $\unit{400}{mm}$ vacuum jacket. The $\unit{250}{mm}$ pipe itself contained the cable pipe ($\unit{100}{mm}$ diameter) and five thinner pipes. The following components contributed to the \Rn\ emanation signal of sample \#52: the inner surface of the $\unit{250}{mm}$ pipe, the outer surface of the cable pipe, as well as the inner and outer surfaces of the included thin pipes. This means the cables and the inner surface of the cable pipe were not included in this measurement. All stainless-steel pipes were electro-polished, except for the bellows. The cryogenic pipe could not be measured separately. Instead, its \Rn\ emanation rate was obtained by measuring it simultaneously with the inner vessel of the cryostat. By subtracting the known result of the latter one (\#48), we inferred a \Rn\ emanation rate of $(\unit{9.4 \pm1.0)}{mBq}$ for it.

Finally, we were interested in the \Rn\ emanation rate of the TPC. From now on we will use the term `inner detector volume' for all subsystems illustrated in Figure \ref{fig:xenon_system} except the purification system. The TPC contribution could be obtained by subtracting the signal of the inner detector volume after and before TPC installation. The latter one was found to be  ($18.4 \pm1.0)$~mBq (see Table \ref{tab:subsystems}) by summing up the contributions from sample \#48 to \#52 and adding the heat exchangers (\#13) and the heater (\#14).
%
Due to the size and complexity of the inner detector volume with the TPC, only a fraction of the carrier gas could be extracted and the final result was obtained by up-scaling as described in section \ref{ema-technique}.  Up-scaling is only appropriate in case of a homogeneous \Rn\ distribution in the carrier gas. Thus, we thoroughly mixed the sample gas immediately before the extraction by adding \Rn-free carrier gas. Moreover, we extracted from various ports to ensure that locations with possibly different \Rn\ activity concentrations were averaged out. More details on the procedure can be found in \cite{stefan_phd}. From the obtained result of $(\unit{19.3 \pm 2.1)}{mBq}$ we calculated the activity of the TPC alone (\#54) which turned out to be compatible with zero. The result could be converted to an upper limit of 4.4~mBq at $90~\%$ C.L. This was in agreement with the known \Rn\ sources of the TPC as quoted in section \ref{components}.

\begin{figure*}[ht!]
\centering
\includegraphics[height=8cm]{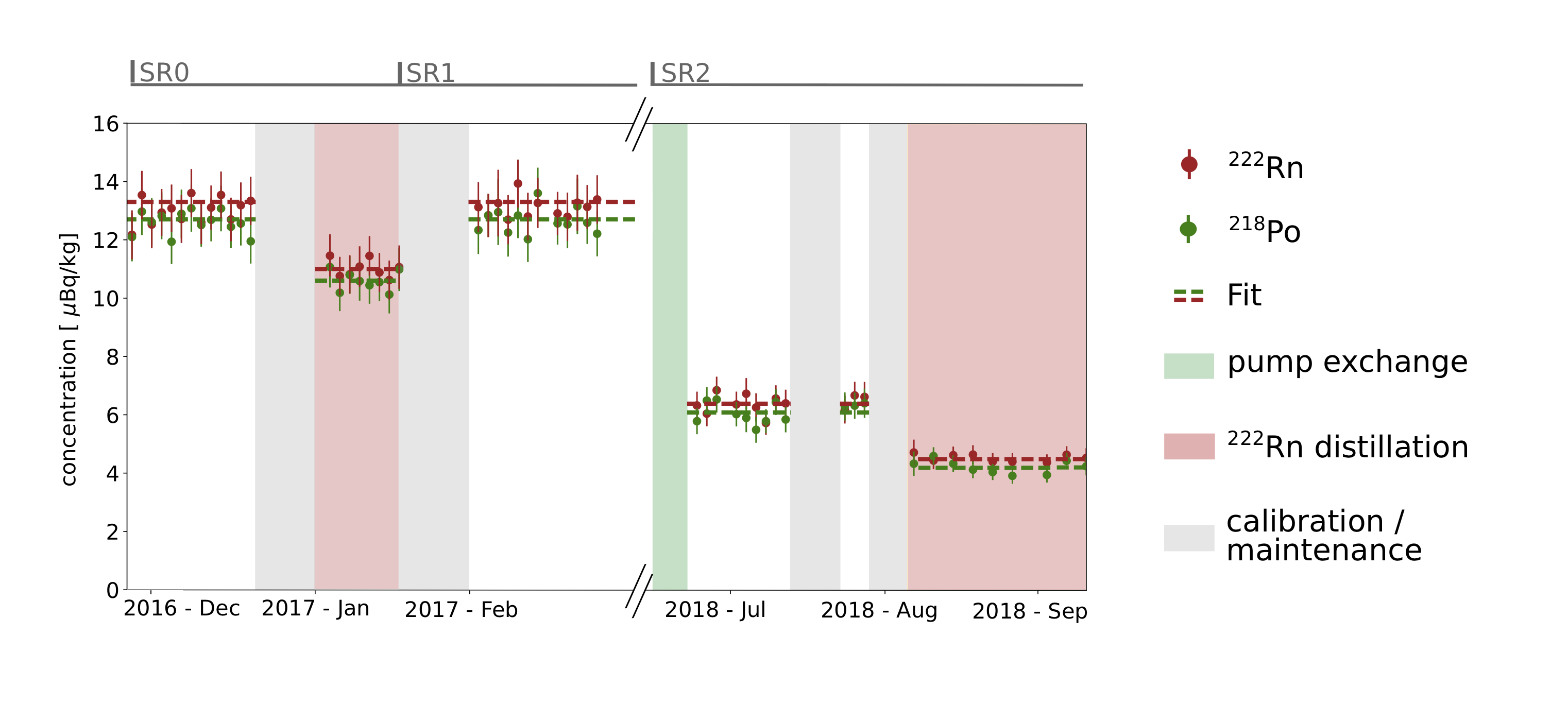}
\caption{The activity concentration evolution of $^{222}$Rn and $^{218}$Po during XENON1T data-taking. In science run 0 and 1 (SR0 and SR1) the activity concentrations were stable over the entire time and we only show the initial period here. Xenon distillation campaigns to remove $^{222}$Rn as well as the exchange of the recirculation pumps lead to a reduction of the $^{222}$Rn and $^{218}$Po activity concentration. They gray regions indicate periods of detector calibration or maintenance.}
 \label{fig:radon_rates}
\end{figure*}

\subsection{Overall $^{222}$Rn budget in XENON1T}
\label{sub:rn_budget}

The measurements of the subsystems allowed us to calculate the expected \Rn\ budget for XENON1T. Apart from the inner detector volume (\#53), the xenon was always in contact with the xenon recirculation pumps and the gas purifiers in the purification system. The measurements of their \Rn\ emanation rates were presented in section \ref{XEPUR}. The two gas purifiers together emanated ($1.41 \pm 0.15$)\,mBq (see Table \ref{tab:getter}). During the main dark matter search phases, the so-called science run 0 (SR0 \cite{SR0_paper}) and science run 1 (SR1 \cite{SR1_paper}), three QDrive pumps were used: C204 (after repair, \#20a), C206 (\#22) and C207. Their \Rn\ emanation rates can be taken from Table \ref{tab:QDrive1} except for pump C207, which was not measured. Its signal could be estimated by taking the average of the highest (\#20) and the lowest (\#21) emanation rate measured for the QDrive pumps with an enlarged systematic error to cover both results within the uncertainty range. With that assumption, we obtained ($4.4 \pm 0.9$)\,mBq for pump C207. Thus, the estimated \Rn\ emanation rate of all three pumps together was ($11.3 \pm 0.9$)\,mBq.

Summing up all components, we obtained a total \Rn\ budget for XENON1T of ($32.0 \pm 2.3$)\,mBq. Under the assumption of a homogeneous radon distribution, we expected a \Rn\ activity concentration of $(\unit{10.0 \pm 0.7)}{\upmu Bq/kg}$ with a total xenon mass of $\unit{3.2}{t}$ in XENON1T meeting the design goal.  The pie chart in Figure \ref{fig:piechart} presents the relative contribution of all components. The dominant elements were the QDrive pumps and the cryogenic pipe which together accounted for about two-thirds of the entire \Rn\ budget.

$\upalpha$-decays of $^{222}$Rn ($\unit{5.5}{MeV}$) and $^{218}$Po ($\unit{6.0}{MeV}$) in the TPC could be selected by their reconstructed energy \cite{radon_xenon100, sander_phd, natascha_phd}. The $\sim\unit{1}{\%}$ energy resolution of the detector at these high energies was sufficient to allow for a clear separation of the two $\upalpha$-peaks. Other background sources in that energy range were subdominant and could be ignored. Therefore, the $\upalpha$-analysis represented a reliable way to measure the \Rn\ and $^{218}$Po activity concentrations in the detector. In Figure \ref{fig:radon_rates}, we show the $\upalpha$-rate evolution during the science runs. An average activity concentrations of $(13.3 \pm 0.5)\,\upmu$Bq/kg and $(12.7 \pm 0.5)\,\upmu$Bq/kg for $^{222}$Rn and $^{218}$Po, respectively, was found for SR0 and SR1, excluding the distillation period for radon removal discussed in section \ref{sub:reduction}. 
$^{218}$Po is often positively charged after its creation \cite{exo_ions}. Thus, ion drift as well as convective motion may transport $^{218}$Po out of the analysis volume and cause its deposition on TPC surfaces. This may explain the slightly lower observed $^{218}$Po activity concentration with respect to $^{222}$Rn.

The discrepancy between the $(13.3 \pm 0.5)\,\upmu$Bq/kg deduced from $\upalpha$-analysis and the expectation of \linebreak ($10.0 \pm 0.7$)\,$\upmu$Bq/kg from emanation measurements corresponds to (10.6 $\pm$ 2.8)\,mBq inside $\unit{3.2}{t}$ of liquid xenon.  A possible explanation for the discrepancy was the \Rn\ release from the QDrive recirculation pumps.  As most samples in this work, they were measured at room temperature.  However, during operation, they heated up and the diffusion-driven \Rn\ emanation could have been enhanced at elevated temperature. In addition, it could have been that the unmeasured QDrive pump C207 emanated more than expected. 

\subsection{Reduction of $^{222}$Rn}
\label{sub:reduction}

There are several possibilities to further reduce the $^{222}$Rn budget. The best option is to remove \Ra, the mother nucleus of \Rn, from the experiment. This was achieved by replacing the QDrive pumps in science run 2 (SR2). They were exchanged for the magnetically coupled piston pump described in section \ref{recirculation-pumps} \cite{muenster-pump}. The decrease of the $^{222}$Rn activity concentration in liquid xenon before and after the pump exchange is shown in Figure \ref{fig:radon_rates}. It corresponded to an absolute reduction of $(19.2 \pm 1.0)\,$mBq in $\unit{3.2}{t}$ of xenon. From the room temperature emanation measurements, presented in Table \ref{tab:QDrive1}, one would have expected a reduction of only ($11.0 \pm 0.9$)\,mBq\footnote{The sum of the signals of the three replaced QDrive pumps minus the signal of the newly mounted magentically coupled piston pump.}. The observed difference of $(8.2 \pm 1.3)$\,mBq agreed within the uncertainty with the difference between \Rn\ emanation measurements and $\upalpha$-analysis (see section \ref{sub:rn_budget}). This supported the hypothesis that the \Rn\ emanation rate of the QDrive pumps was larger than estimated. By exchanging the recirculation pumps, the largest $^{222}$Rn source was successfully removed from the experiment yielding a reduced background level \cite{lowE_paper}. This is very promising for XENONnT, where magnetically coupled piston pumps will be employed.

Another possibility for \Rn\ reduction is an online radon removal system in the xenon purification loop, which separates xenon from radon and retains the latter until its disintegration.  Such a radon removal system based on cryogenic distillation was pioneered by the XENON collaboration \cite{xenon_rn_dist, stefan_boiloff}. In XENON1T, we realized radon removal by employing the cryogenic distillation system built for krypton removal from xenon and operating it in reverse direction \cite{murra_phd}.  As shown in Figure \ref{fig:radon_rates}, the distillation led to a $^{222}$Rn reduction of $\sim\unit{20}{\%}$ during SR0, although only a small fraction of the recirculation flow was distilled. A compatible absolute \Rn\ reduction was achieved in a second xenon distillation run performed in SR2, after the pump exchange. The finally accomplished \Rn\ activity concentration in XENON1T was $\unit{(4.5\pm0.1)}{\upmu Bq/kg}$.

As a consequence of the promising results, the XENON collaboration realized two new purification systems for \linebreak XENONnT which complement the existing ones and sig\-ni\-fi\-cant\-ly improve the xenon purity. The first one is a novel li\-quid purification system which is able to produce and maintain ultra-pure liquid xenon at a very fast flow rate. The se\-cond one is a dedicated distillation column which was designed \cite{murra_phd} and built for radon removal and which takes advantage of the large flow rate enabled by the liquid purification system. 

The impact of the new radon removal system on the experiment's background can be further maximized if \Rn\ is flushed out of the detector before it enters the TPC. A detailed mapping of the \Rn\ sources in XENON1T was obtained by the \Rn\ emanation measurements of the various subsystems, presented in this work. Thus, a targeted xenon flow pattern optimization could be studied. Such a flow pattern optimization with respect to the radon removal system will be applied in the XENONnT experiment \cite{XENONnT_sensitivity}.


\section{Summary and outlook}
\label{summary}

The background rate of current xenon dark matter detectors is dominated by \Rn-induced events and it is expected that \Rn\ daughters will remain an essential background component in future experiments. Therefore, \Rn\ emanation measurements become increasingly important and provide complementary information to the bulk radioactivity screening efforts. In this article, we presented the results of a comprehensive material screening campaign for \Rn\ emanation carried out for the XENON1T experiment, using state-of-the-art counting techniques. We selected construction materials with the lowest possible \Rn\ emanation rate and we were able to identify and locate the remaining \Rn\ sources in the experiment.

The predicted activity concentration from these measurements was in agreement with the target \Rn\ activity concentration of $\unit{10}{\upmu Bq/kg}$ in $\unit{3.2}{t}$ of xenon \cite{xenon_physics_reach}. Results from an $\upalpha$-analysis of the XENON1T data were about $30~\%$ higher than the prediction.  The discrepancy could be understood by an underestimation of the recirculation pump's \Rn\ emanation rate.  With the exact knowledge of the distribution of \Rn\ sources in XENON1T, it became possible to selectively eliminate problematic items. The ultimately measured \Rn\ activity concentration of $\unit{(4.5\pm0.1)}{\upmu Bq/kg}$ is the lowest ever achieved in a xenon dark matter experiment. Significant improvements are possible in XENONnT and further projects, for instance, by continuous xenon distillation.

\section{Acknowledgements}

We gratefully acknowledge support from the National Science Foundation, Swiss National Science Foundation, German Ministry for Education and Research, Max Planck Ge\-sellschaft, Deutsche Forschungsgemeinschaft, Netherlands Organisation for Scientific Research (NWO), Weizmann Institute of Science, ISF, Fundacao para a Ciencia e a Tecnologia, R\'egion des Pays de la Loire, Knut and Alice Wallenberg Foundation, Kavli Foundation, JSPS Kakenhi in Japan and Istituto Nazionale di Fisica Nucleare. This project has received funding or support from the European Union's Horizon 2020 research and innovation programme under the Ma\-rie Sklodowska-Curie Grant Agreements No. 690575 and No. 674896, respectively. Data processing is performed using infrastructures from the Open Science Grid, the European Grid Initiative and the Dutch national e-infrastructure with the support of SURF Cooperative. We are grateful to Laboratori Nazionali del Gran Sasso for hosting and supporting the XENON project.


\begin{thebibliography}{99}

\bibitem{planck_cosmo_constrains}
Planck Collaboration \emph{Planck 2018 results. VI. Cosmological parameters}, 	arXiv:1807.06209 (2018).
\bibitem{instrument_xenon}
E. Aprile et al. (XENON Collaboration), \emph{The XENON1T Dark Matter Experiment}, Eur. Phys. J. C77:881 (2017).
\bibitem{Xe10}
E. Aprile et al. (XENON10 Collaboration), \emph{Design and Performance of the XENON10 Dark Matter Experiment}, Astropart.Phys.34:679-698 (2011).
\bibitem{Xe100}
E. Aprile et al. (XENON100 Collaboration), \emph{The XENON100 Dark Matter Experiment}, Astropart. Phys. 35, 573-590 (2012).
\bibitem{strigari}
L. E. Strigari, \emph{Galactic Searches for Dark Matter}, Phys. Rep. 531, 1 (2013).
\bibitem{nexo}
nEXO Collaboration, \emph{nEXO Pre-Conceptual Design Report}, arXiv:1805.11142 (2018).
\bibitem{LZ}
B.J. Mount et al. (LZ Collaboration), \emph{LUX-ZEPLIN (LZ) Technical Design Report}, arXiv:1703.09144 (2017).
\bibitem{pandaX}
Li Zhao et al. (PandaX Collaboration), \emph{PandaX: A deep underground dark matter search experiment in China using liquid xenon}, Mod. Phys. Lett. A33 (2018) no.30, 1830013  (2018).
\bibitem{xenon_physics_reach}
E. Aprile et al. (XENON Collaboration), \emph{Physics reach of the XENON1T dark matter experiment}, J. Cos. Astrop. Phys. 04:027 (2016)
\bibitem{gamma_screening_paper}
E. Aprile et al. (XENON Collaboration), \emph{Material radioassay and selection for the XENON1T dark matter experiment}, Eur. Phys. J. C77:890 (2017).
\bibitem{balek}
V. Balek, \emph{Emanation thermal analysis}, Thermochim. Acta 222 (1978) 1.
\bibitem{recoil-reference}
International Atomic Energy Agency \emph{Nuclear Data Services}, https://www-nds.iaea.org/.
\bibitem{xe-diffusion}
R. Nagasaki and S. Kawasaki, \emph{Diffusion of xenon in metals}, J. Nucl. Materials 19(1):90-92 (1966).
\bibitem{GERDA_welds}
G. Zuzel and H. Simgen, \emph{High sensitivity radon emanation measurements}, Appl. Radiat. Isot. 67(5):889-893 (2009).
\bibitem{Wink-paper}
R. Wink et al., \emph{The miniaturized proportional counter HD-2(Fe)/(Si) for the GALLEX solar neutrino experiment}, Nucl. Inst. Meth. A 329:3, 541 (1993).
\bibitem{gas-line} H. Simgen et al., \emph{Highly sensitive measurements of radioactive noble gas nuclides in the Borexino solar neutrino experiment}, App. Rad. Isot. 61, 213 (2004).
\bibitem{radon_monitor}
J. Kiko, \emph{ Detector for $^{222}$Rn measurements in air at 1 mBq/m$^{3}$ level}, Nucl. Instrum. Methods A, 460 (2001).
\bibitem{stefan_phd}
S. A. Bruenner, \emph{Mitigation of $^{222}$Rn induced background in the XENON1T dark matter experiment}, PhD thesis, University of Heidelberg, https://doi.org/10.11588/heidok.00023261 (2017).
\bibitem{kr_distillation}
E. Aprile et al. (XENON Collaboration),  \emph{Removing krypton from xenon by cryogenic distillation to the ppq level}, Eur. Phys. J. C 77, 275 (2017).
\bibitem{SimgenBX}
H. Simgen et al., \emph{Analysis of radioactive trace impurities with $\upmu$Bq-sensitivity in Borexino}, Int. J. Mod. Phys. A 29,16 1442009 (2014).
\bibitem{xenon_rn_dist}
E. Aprile et al. (XENON Collaboration), \emph{Online $^{222}$Rn removal by cryogenic distillation in the XENON100 experiment}, Eur. Phys. J. C77 (2017).
\bibitem{qdrive-pump}
Chart Industries, https://www.chartindustries.com/
\bibitem{exo200-pump}
F. LePort et al., \emph{A magnetically driven piston pump for ultra-clean applications}, Rev. Sci. Instr. 82, 105114 (2011).
\bibitem{muenster-pump}
E. Brown et al.,  \emph{Magnetically-coupled piston pump for high-purity gas applications}, Eur. Phys. J., C78:604 (2018).
\bibitem{PMT-paper}
E. Aprile et al. (XENON Collaboration), \emph{Lowering the radioactivity of the photomultiplier tubes for the XENON1T dark matter experiment}, Eur. Phys. J. C75:546 (2015).
\bibitem{hamamatsu}
Hamamatsu Photonics, \emph{Handbook Photomultiplier tubes Basics and Applications}, https://www.hamamatsu.com 
\bibitem{220Rn-source}
R.F. Lang et al., \emph{A Rn-220 source for the calibration of low-background experiments}, JINST 11 P04004 (2016).
\bibitem{natascha_master}
N. Rupp \emph{On the detection of $^{222}$Rn with miniaturized propotional counters: background, sensitivity studies and results for XENON1T}, Master thesis, University of Heidelberg, http://hdl.handle.net/11858/00-001M-0000-0027-797D-D (2015).
\bibitem{XENONnT_sensitivity}
E. Aprile et al. (XENON Collaboration), \emph{Projected WIMP Sensitivity of the XENONnT Dark Matter Experiment}, arXiv:2007.08796 (2020).
\bibitem{SR0_paper}
E. Aprile et al. (XENON Collaboration), \emph{First Dark Matter Search Results from the XENON1T Experiment}, Phys. Rev. Lett. 119, 181301 (2017).
\bibitem{SR1_paper}
E. Aprile et al. (XENON Collaboration), \emph{Dark Matter Search Results from a One Ton-Year Exposure of XENON1T}, Phys. Rev. Lett. 121, 111302 (2018).
\bibitem{radon_xenon100}
E. Aprile et al. (XENON Collaboration),  \emph{Intrinsic backgrounds from Rn and Kr in the XENON100 experiment}, Eur. Phys. J. C78: 132 (2018).
\bibitem{sander_phd}
P.A. Breur,  \emph{Backgrounds in XENON1T} Phd thesis, University of Amsterdam, https://hdl.handle.net/11245.1/7bed5cdb-57f2-4c13-b09c-407936a669de (2019).
\bibitem{natascha_phd}
N. Rupp,  \emph{PhD thesis in preparation} (2020).
\bibitem{exo_ions}
EXO-200 Collaboration,  \emph{Measurements of the ion fraction and mobility of alpha and beta decay products in liquid xenon using EXO-200}, Phys. Rev. C 92, 045504 (2015).
\bibitem{lowE_paper}
E. Aprile et al. (XENON Collaboration), \emph{Observation of Excess Electronic Recoil Events in XENON1T},  arXiv:2006.09721v2 (2020).
\bibitem{stefan_boiloff}
S. A. Bruenner et al., \emph{Radon depletion in xenon boil-off gas}, Eur. Phys. J., C77:143, (2017).
\bibitem{murra_phd}
M. Murra, \emph{Intrinsic background reduction by cryogenic distillation for the XENON1T dark matter experiment}, PhD thesis, WWU Münster, (2019).
\end{thebibliography}
\end{document}